\documentstyle[11pt]{article}
\input psfig

\def\kms{km s$^{-1}$}
\def\ha{H$\alpha$}
\def\Msun{M_{\odot \hskip-5.2pt \bullet}}

\reversemarginpar

\begin{document}

\title{Central Rotation Curves of Galaxies}
\author{Yoshiaki SOFUE \\
Institute of Astronomy, University of Tokyo, Mitaka,
Tokyo 181-0015, Japan: sofue@ioa.s.u-tokyo.ac.jp}
\maketitle
\begin{abstract}
We emphasize the use of high-resolution CO line observations to 
derive central rotation curves of galaxies.
We present an example for high-resolution interferometer observations of
NGC 3079, and discuss the PV diagram and derived rotation curve.
The CO central rotation curves are combined with the outer curves from
\ha\ and HI-line observations to obtain total RC.
We show that well resolved RCs for massive galaxies 
generally start from non-zero
velocities at the nucleus.
\end{abstract}

\section{Introduction}

The CO molecular lines are useful to derive
accurate rotation curves in the central regions of spiral galaxies,
because of the high concentration in the center as well as for
negligible extinction (Sofue 1996, Sofue et al 1997, 1998, 1999).
Recent high-dynamic-range CCD spectroscopy in optical lines
also make it possible to obtain high accuracy rotation curves in the central
regions (Rubin et al 1997; Bertola et al 1998).
However, in general,  optical lines suffer from significant 
extinction by the central dusty disks, which is particularly signficant
for highly-inclined and 
edge-on galaxies.  Hence, the CO lines will be the most appropriate 
tool to investigate the central kinematics of spiral galaxies, 
if the angular resolution is sufficiently high.
In this paper, we present recent high-resolution interferometer observations
of CO lines, and discuss the general properties of central rotation kinematics
based on the high-accuracy rotation curves.

\begin{figure}
\psfig{figure=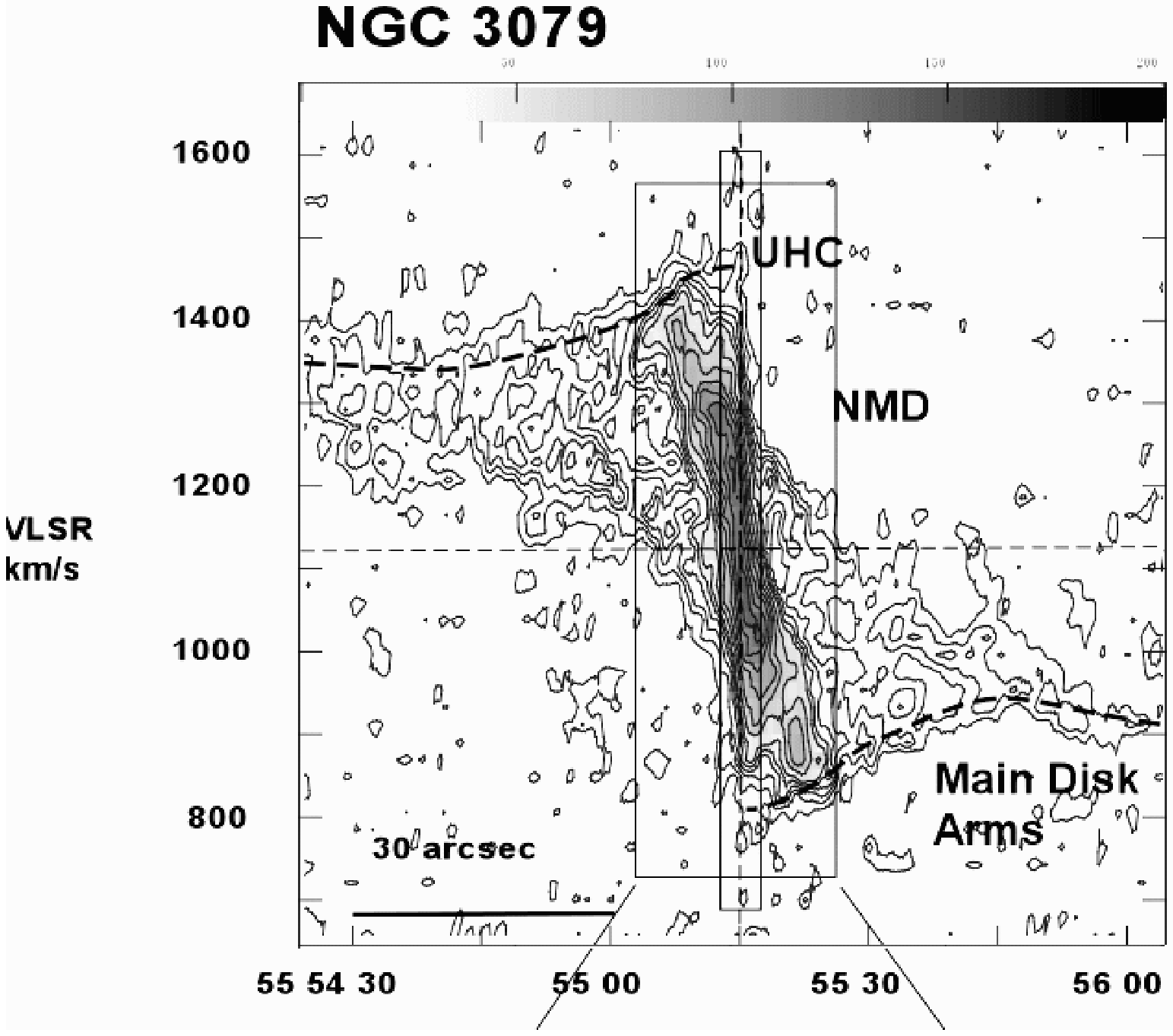,width=10cm}
\psfig{figure=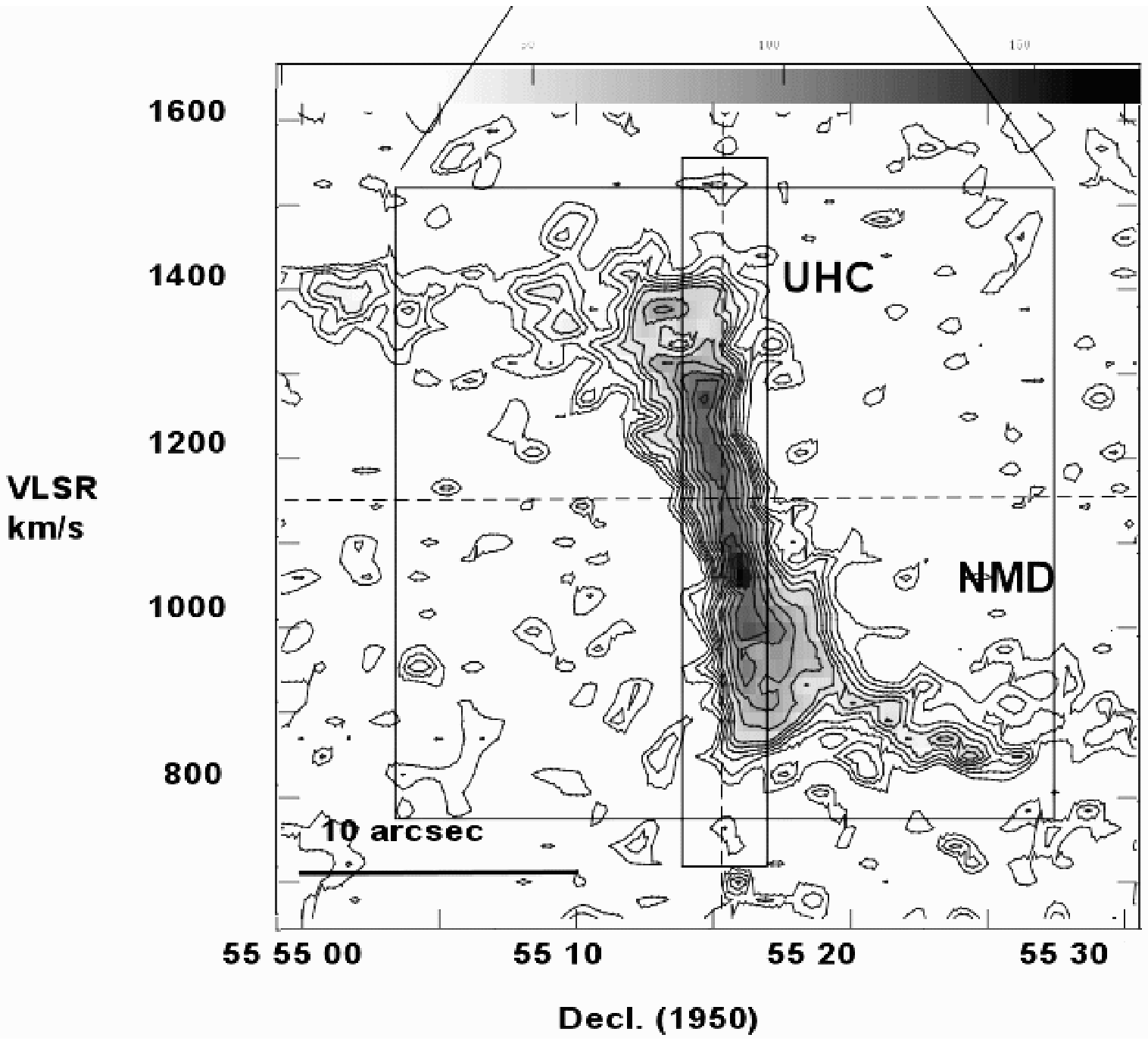,width=10cm}
Fig. 1. Position-velocity diagram along the major axis of NGC 3079 in
CO J=1-0 line at resolutions of $1''6 \times 1.4''$ (top) and 
$1''.2 \times 1''.1$ (bottom). 
Note that the rotation velocity does not decline to zero
at the nucleus, but remains always at finite value.
UHC stands for ultra-high-density molecular core, and NMD for
nuclear molecular disk. $1''$ corresponds to 75 pc.
\end{figure}

\section{High-resolution CO Observations of NGC 3079}

In order to see if the central steep rise, or more likely  non-zero
start of the velocity, is indeed the case at higher resolutions
than the current observations,
we have performed interferometer observations at Nobeyama 
in the CO (J=1-0) line of nearby CO-rich galaxies.
Here, we present an example for the edge-on galaxy NGC 3079, which exhibits
a very dense central molecular core with various nuclear activities
(Sofue et al 2000).
The  $^{12}$CO (1 - 0) observations of NGC3079 were made in January to April
2000 using the 7-element mm-wave interferometer at Nobeyama, 
which consisted of the 6-element mm-wave array in A configuration
linked with the 45-m telescope. We also obtained C and D-compact array
observations, and all UV data were combined.
Fig. 1 shows the obtained PV (position-velocity) diagrams.

The CO line intensity distribution in NGC 3079 is summarized as follows:
(a) An ultra-high-density compact and massive molecular core (UHC) of radius
125 pc and molecular mass of $3 \times 10^8\Msun$  is detected
at the nucleus.
(b) The core is embedded in a warped nuclear molecular disk (NMD) 
of radius 900 pc, and the disk has two spiral arms.
(c) An outer disk extending for more than 2 kpc along
the  major axis.

\begin{figure}
\psfig{figure=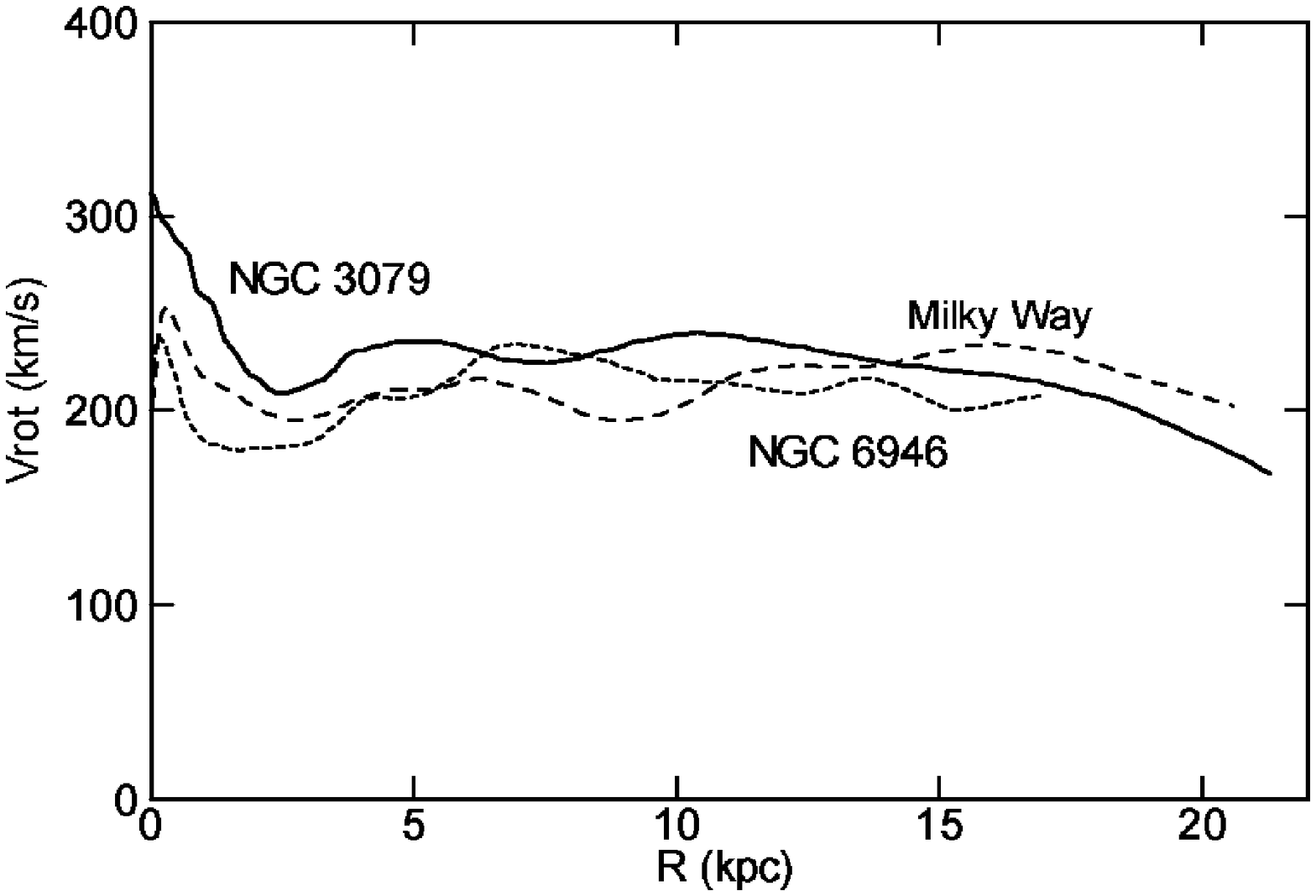,width=6.5cm}
Fig. 2 (left). Rotation curve of NGC 3079, compared with that of the Milky Way and NGC 6946.
\end{figure}

\begin{figure}
\psfig{figure=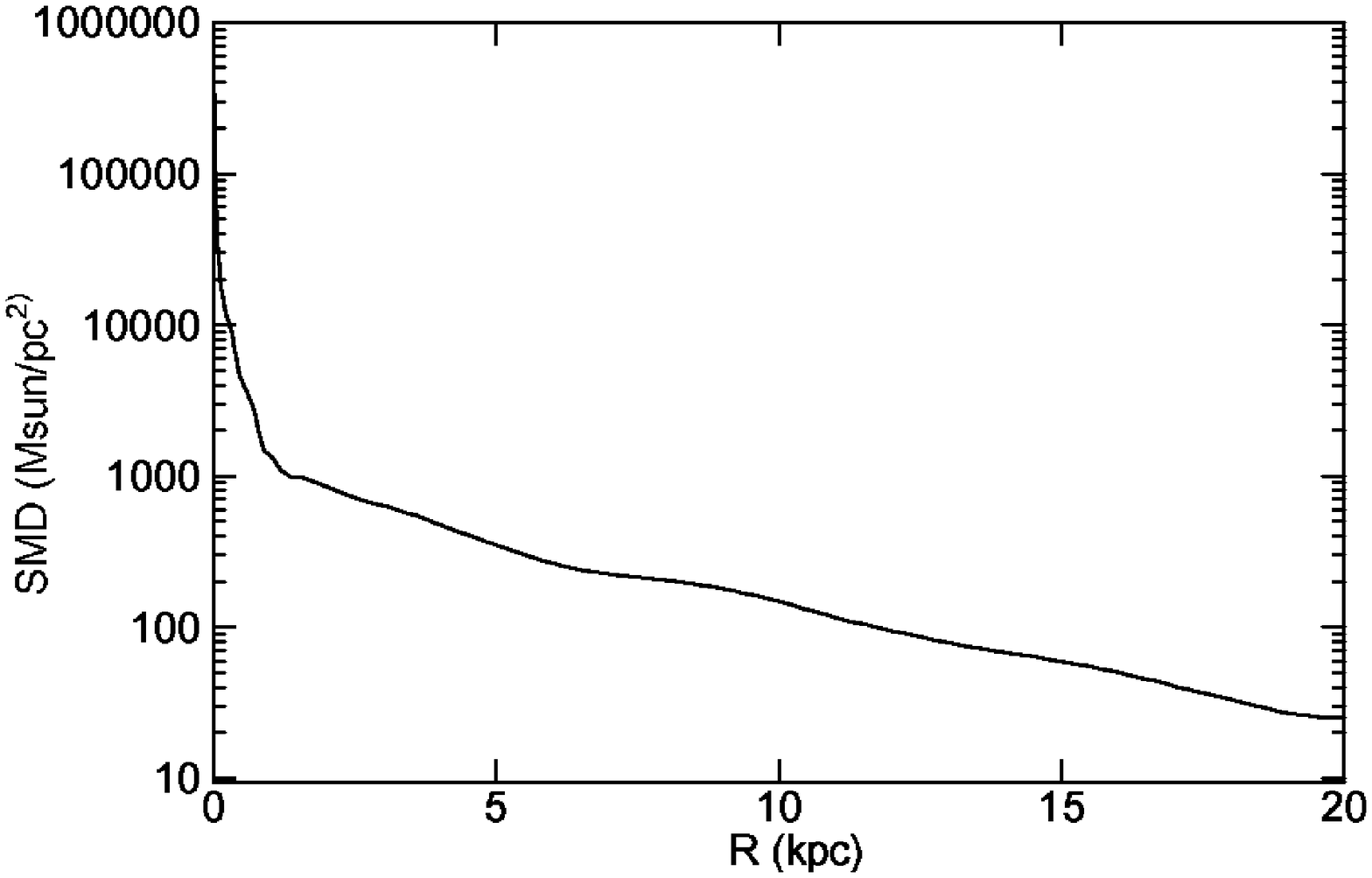,width=6.6cm}
Fig. 3 (right). SMD (Surface mass density) distribution in NGC 3079.
\end{figure}

\section{Position-Velocity Diagrams and Rotation Curve}

Fig. 1 shows the obtained PV diagrams for the central regions of
NGC 3079, where the central UHC and NMD are resolved out both in
the velocity and space.
The UHC shows up as an intense PV ridge near the center.
The rotation velocity of the molecular core increases
toward the nucleus, and the velocity does not decline to zero at the center,
indicating that the rotation curve starts from a finite value already at
the center with at about 300 \kms\ or greater.
The warped nuclear disk shows up as an inclined ridge in
the  PV diagram, representing two-armed spiral arms.
The radius of this disk component 
is $\pm12''$, or the total radius is about $\pm900$ pc.
The main disk of the galaxy in the PV diagram shows up as two
fainter ridges with smaller relative velocities,
bifurcating from the main ridge of the nuclear disk.
These bifurcated ridges show roughly rigid rotation, but  at
slower velocities, which represents foreground/background spiral arms.
We also notice  a velocity displacement between the two arm-like features,
which may  represent non-circular motion driven by the spiral density waves.

Using the PV diagram, we derived a central rotation curve, and combined it with
the existing data (e.g. those from Irwin and Seaquest 1991)
to obtain total rotation curve as shown in Fig. 2.
The rotation velocity starts from a finite value of about 300 \kms,
and declines to a first minimum of about 200 \kms\ at 30$''$ (2.5 kpc) radius.
It is then followed by a  broad disk maximum of $V \sim 240$ \kms at 
5 to 10 kpc, and then by a declining outermost  rotation.

Using the rotation curve, we can derive a differential surface mass
density as a function of  radius by applying the method developed by
Takamiya and Sofue (1999).
The surface mass densitu (SMD) increases steeply toward the center,
indicating high density cores with $SMD >10^5 \Msun$ pc$^{-2}$.
Since NGC 3079 is an Sc galaxy with a poor central bulge, the
high mass density could infer a concentration of invisible (dark) 
mass in the central region.

\begin{figure}
\psfig{figure=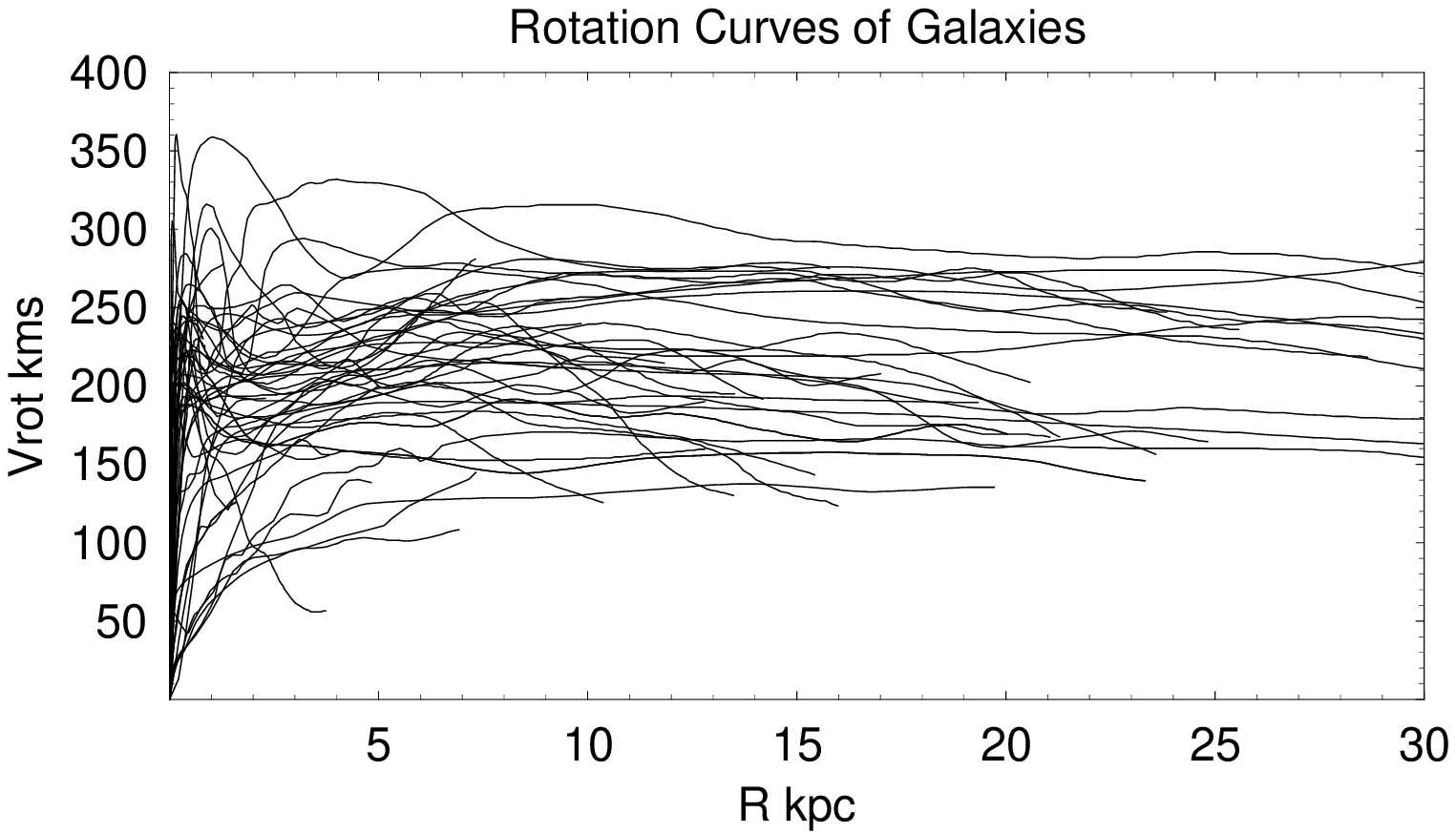,height=6cm}
Fig. 4. High-accuracy rotation curves  of Sb, Sc, SBb
and SBc galaxies.
\end{figure}

\section{Universal Properties of Rotation Curves}

In Fig. 4 we reproduce well-sampled rotation curves obtained by
combining CO, CCD H$\alpha$, and HI observations from  our current
study  (Sofue et al 1999).
From Figs. 2, 3 and 4, particularly from the case for the highest-quality
rotation curves for NGC 3079, we may summarize the universal properties of  
rotation curves as follows.

(1)Massive galaxies generally show very steep rise of rotation in the 
central region.
Mostly likely, the rotation velocity starts from a finite value 
at the center, indicating a massive core at the nucleus.
(2) Small-mass galaxies with slowe rotation velocities, however,
 tend to show more gentle rise.
(3) The central rotation is followed by a central peak and/or 
shoulder corresponding to the bulge.
(3) RC is then followed by a road maximum in the disk.
(4) The outer RC is flat, and sometimes declining toward the edge.

\vskip 2mm 
\parskip=0pt 
\def\r{\hangindent=1pc \noindent}

\noindent{References}

\r Bertola, F., Cappellari, M., Funes, J.G., et al 1998, ApJ, 509, L93.

\r Irwin, J. A. and Seaquest, E. R. 1991 ApJ 371, 111.

\r Rubin, V., Kenney, J.D.P., Young, J.S. 1997 AJ, 113, 1250.

\r Sofue, Y. 1996, ApJ, 458, 120

\r Sofue, Y.,  Tutui, Y., Honma, M., and Tomita, A., 1997, AJ, 114, 2428

\r Sofue, Y., Tomita, A.,  Honma, M., et al. 1998, PASJ 50, 427.

\r Sofue, Y., Koda, J., Kohno, K., et al.  2000, submimtted to ApJ. L.

\r Sofue, Y., Tutui, Y., Honma, M., et al.  1999, ApJ, 523

\r Takamiya, T., and Sofue, Y. 1999, to appear in ApJ.

\end{document}